\documentclass[11pt,a4paper]{article}

\AtBeginDocument{%
  }

\usepackage[utf8]{inputenc}
\usepackage{graphicx}     
\usepackage{amsmath}      
\usepackage{lmodern}
\usepackage{authblk}
\usepackage{eurosym}
\usepackage{paralist}
\usepackage{mdframed}
\usepackage{blindtext}
\usepackage{soul}
\usepackage{comment}
\usepackage[TS1,T1]{fontenc}

\usepackage{enumitem}
\usepackage{multirow}
\usepackage[labelfont=bf]{caption}

\usepackage{csquotes}

\usepackage{array}
\usepackage{longtable}

\usepackage{enumitem}

\usepackage{hyperref}
\usepackage{xurl}

\usepackage[
  a4paper,
  left=2cm,
  right=2cm
]{geometry}

\usepackage[nomain,acronym]{glossaries}
\makeglossaries

\newcommand{\abkuerzung}[2]{\newacronym{#1}{#1}{#2}}

\abkuerzung{AADL}{Architecture Analysis \& Design Language}

\abkuerzung{AAS}{Asset Administration Shell}

\abkuerzung{ADL}{architecture description language}

\abkuerzung{ADR}{Action Design Research}

\abkuerzung{API}{application programming interface}

\abkuerzung{AS}{abstract syntax}

\abkuerzung{AST}{abstract syntax tree}

\abkuerzung{CBSE}{component-based software engineering}

\abkuerzung{CD}{Class Diagram}

\abkuerzung{CFG}{context-free grammar}

\abkuerzung{CoCo}{context condition}

\abkuerzung{CS}{concrete syntax}

\abkuerzung{CPS}{cyber-physical system}

\abkuerzung{DoE}{design of experiments}

\abkuerzung{DSL}{domain-specific language}

\abkuerzung{DS}{Digital Shadow}

\abkuerzung{DSRM}{Digital Shadow Reference Model}

\abkuerzung{DT}{Digital Twin}

\abkuerzung{FSM}{finite-state machine}

\abkuerzung{GPL}{general-purpose programming language}

\abkuerzung{KR}{knowledge representation}

\abkuerzung{LCDP}{low-code development platform}

\abkuerzung{DPL}{DSL product line}

\abkuerzung{M2M}{model-to-model}

\abkuerzung{M2T}{model-to-text}

\abkuerzung{MBSE}{model-based systems engineering}

\abkuerzung{MDE}{model-driven engineering}

\abkuerzung{MDSE}{model-driven systems engineering}

\abkuerzung{OCL}{Object Constraint Language}

\abkuerzung{SC}{statechart}

\abkuerzung{Sem}{semantics}

\abkuerzung{SME}{small and medium-sized enterprise}

\abkuerzung{SLE}{software language engineering}

\abkuerzung{SOS}{Structural Operational Semantics}

\abkuerzung{SPL}{software product line}

\abkuerzung{SysML}{Systems Modeling Language}

\abkuerzung{TS}{technological space}

\abkuerzung{TSCA}{time-synchronous channel automata}

\abkuerzung{TSPA}{time-synchronous port automata}

\abkuerzung{UML}{Unified Modeling Language}

\usepackage{xspace}
\usepackage{ifdraft}

\makeatletter
\newcommand*{\etc}{%
  \@ifnextchar{.}%
  {\textit{etc}}%
  {\textit{etc.}\@\xspace}%
}
\makeatother

\definecolor{se-green}{RGB}{0,128,0}
\definecolor{se-blue} {RGB}{0,0,204}

\newcommand{\spic}[2]{\includegraphics*[width=#1\textwidth]{#2}}

\definecolor{MyBoxText}{RGB}{255,255,255}
\definecolor{MyBoxBG}{RGB}{13,50,153}

\title{Towards a Platform for Mastering Personal Sovereignty}

\author[1]{Judith Michael}
\author[2]{Bernhard Rumpe}
\author[3]{Antonio Bucchiarone}
\author[4]{Dominik Bork}

\affil[1]{University of Regensburg, Regensburg, Germany\\
\texttt{judith.michael@ur.de}}
\affil[2]{RWTH Aachen University, Aachen, Germany\\
\texttt{rumpe@se-rwth.de}}
\affil[3]{SWEN/DISIM, Universit\`a degli Studi dell'Aquila, L'Aquila, Italy\\
\texttt{antonio.bucchiarone@univaq.it}}
\affil[4]{TU Wien, Business Informatics Group, Vienna, Austria\\
\texttt{dominik.bork@tuwien.ac.at}}

\newcommand{\keywords}[1]{%
  \par\noindent\textbf{Keywords:} #1}

\begin{document}

\emergencystretch 3em

\maketitle

\begin{abstract}

The ongoing digital transformation of work, administration, health, mobility, and social interaction is profoundly reshaping everyday life, steadily shifting control from individuals to large platform providers. 
Although data is often labeled the “gold of the 21st century”, its real value is realized through services that access, combine, and exploit it. 
Today, individuals have little sovereignty: life events (e.g., changing an address, insurance, job, or marital status) require fragmented, repetitive interactions across numerous systems, leaving users overwhelmed rather than empowered. 
We argue for a fundamental rethinking of this situation and propose a virtual, trustworthy platform, called "\texttt{MyVirtualME}" that acts on behalf of the human individual towards companies, administrations, and other actors. 
This platform centers services, data, permissions, and data usage around humans, not foreign actors, allowing genuine control and transparency.
Our vision goes beyond a pure data focus: supported by trusted services ranging from basic notifications to intelligent assistants, \texttt{MyVirtualME} aims to reduce digital bureaucracy, improve and even automate interactions, and support informed oversight of an individual's health, financial, and administrative situation. 
We firmly believe that a human-centered \texttt{MyVirtualME} is essential for restoring personal sovereignty in our digital future.
\end{abstract}


\section{Turning our ways of interaction}\label{sec:motivation}





Everyday life unfolds as a continuous flow of situations in which a person inhabits multiple roles, often without noticing the connections between them.
In the morning, an individual is a family member and a citizen; throughout the day, the same person becomes a professional collaborating in a digital workplace, a learner improving skills, a patient interacting with healthcare services, and, eventually, a reflective individual reviewing personal choices and boundaries. Despite this continuity of life, today’s digital world fragments these moments into a multitude of isolated applications, identities, and redundant data silos, forcing people to repeatedly re-enter information, manually manage permissions, and adapt to the logic of systems rather than the other way around.

So far, the digital transformation has not been human-centered; instead, organizations have established digital processes grounded in isolated data silos that often prevent efficient digital self-management and human sovereignty.
Think about digital public services in many countries that mirror administrative silos, forcing citizens to repeatedly enter the same data, unnecessarily complex opt-out methods, and various 2-factor authentication apps that fill our smartphones.
This often leads to a digital divide~\cite{warschauer2010digital} 
or digital fatigue~\cite{supriyadi2025impact}
, and even the digitally savvy are losing oversight and thus control.

The concept of \texttt{MyVirtualME} proposes a radically different paradigm. It centers on a persistent, active, and trustworthy platform with a virtual representation of the person that accompanies them throughout the day, maintaining continuity across roles while respecting contextual boundaries.
Rather than a single monolithic profile, \texttt{MyVirtualME} dynamically activates role-specific services for each individual, each governed by explicit rules defined by the person.
The person fully controls their \texttt{MyVirtualME} and can rely on it, allowing it to act on their behalf.
It does not use foreign services; instead, it provides its own services to other actors.


\section{A glance of our future with \texttt{MyVirtualME}}\label{sec:glance}

Consider Andy, a citizen and family member who uses his own \texttt{MyVirtualME} to manage his interactions with civic services, administration, schools, and mobility infrastructures.
When Andy shifts into his professional role, \texttt{MyVirtualME} presents an extended digital posture: selectively exposing credentials, availability, and experience.
Later, in learning or healthcare contexts, \texttt{MyVirtualME} draws on long-term health or learning records, omitting for Andy the need to deal with details while exposing him, e.g., to what he needs to do.
In all those different contexts, Andy’s \texttt{MyVirtualME} shares with external services only what is relevant and authorized, and keeps track of what was shared and for what purpose Andy authorized this data sharing.

Throughout this assistance, \texttt{MyVirtualME} memorizes, reasons about, and orchestrates data rather than merely storing it.
It keeps track of what data exist, where it originates, and how it has been used.
It does not replace human decision-making, but reduces friction by negotiating routine interactions, preparing actions, and explaining its behavior in human-understandable terms.
It acts as Andy's personal advocate, while Andy remains the ultimate authority, capable of reviewing past actions, redefining delegation rules, and reclaiming control at any time.
It avoids that Andy is overwhelmed with irrelevant information, e.g., advertisement of unwanted goods, but helps searching for, e.g., the best next car or furniture, when needed.

From this perspective, \texttt{MyVirtualME} is not simply an intelligent assistant, but also a mediator between Andy and an ever-growing, complex digital ecosystem, restoring user-centeredness and aligning digital services around the continuity of Andy's life, rather than around institutional application or service boundaries.
The value of \texttt{MyVirtualME} emerges precisely in transitions between roles, moments, and contexts where today’s systems do not support people and where data sovereignty, transparency, and agency become essential for a livable digital future.

\begin{figure*}
	\centering
	\spic{0.85}{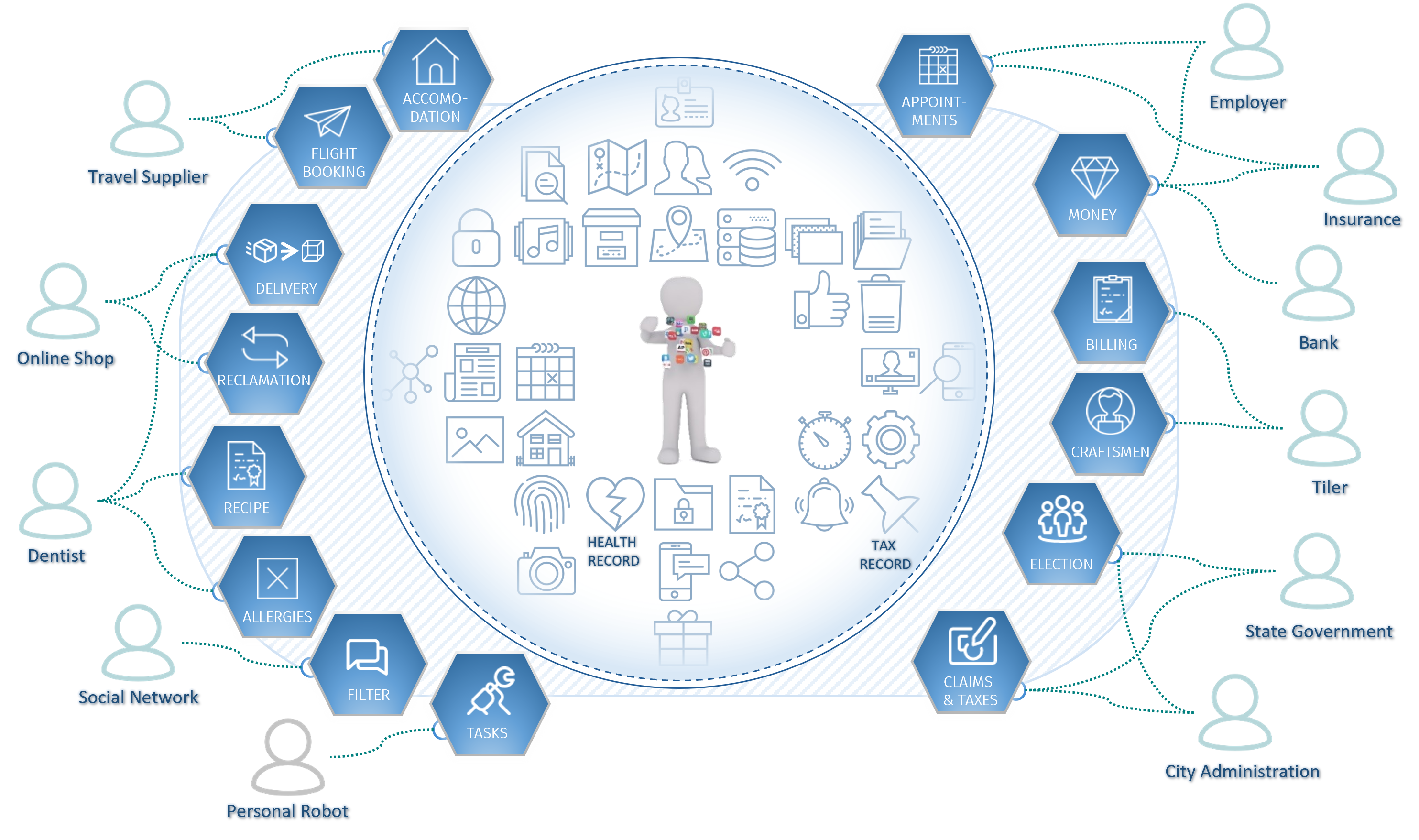}
	\caption{Mastering the personal digital future with the \texttt{MyVirtualME} platform.}
	\label{fig:mvm}
\end{figure*}

Fig.~\ref{fig:mvm} shows a typical single-day choreography of Andy. In the center is Andy’s \texttt{MyVirtualME}, represented as a stable contact that can be seen as a functional digital avatar.
Around Andy and his virtual representation, the day unfolds as softly connected scenes: \textit{home}, \textit{work}, \textit{learning}, \textit{healthcare}, and more.
These scenes are not isolated; the smartness lies in the possibility of overlapping zones, especially when managing and planning financial resources, time, or other relevant goods.

In \texttt{MyVirtualME}, some data remain close to the core, indicating long-term personal memory, such as identity attributes, preferences, values, travel plans, and personal experiences, such as vacation photos, restaurant reviews, or medical treatments.
Other data are mainly aimed at exchange between actors.
Importantly, these streams never flow directly from one actor to another; they always pass through \texttt{MyVirtualME}, reinforcing the idea that the person, not other platforms, is the point of integration, filter, augmentation, and authorization. \textit{\texttt{MyVirtualME} is the conductor of an orchestra and thus itself an active performer that adapts to the rhythm of human life, fully and transparently controlled by the human}.

From \texttt{MyVirtualME}'s viewpoint, there are external services representing external actors positioned at the periphery of the scene: civic systems, learning platforms, and healthcare providers.
They see \texttt{MyVirtualME} as an agent that provides means to negotiate/request/configure access to a bunch of services.
The external actors do not own the data; instead, they interact with \texttt{MyVirtualME} through clear, visible connections only when needed.
\texttt{MyVirtualME} records all interactions, including the reason for contact, what was exchanged, etc. 
Andy controls all of this via a dashboard on top of \texttt{MyVirtualME}.
It allows him to zoom out and see the overall choreography of the day or his future plans, which roles were enacted or will be enacted, which external actors were involved, and how data were used.
Andy can zoom in on a single interaction to understand and adjust it.
Transparency is key: nothing happens invisibly, and while \texttt{MyVirtualME} provides much of the automation, Andy can always adapt the choreography.


\subsection{A more detailed scenario: Health care}\label{sec:scenario1}

The healthcare domain is extremely fragmented, with many different players partially interacting directly and partially expecting patients or treatment recipients to connect with them.
In such a setting, many optimizations are possible with \texttt{MyVirtualME}.

Andy usually sees a variety of doctors with different specialties and visits different hospitals for medical treatment.
They need the exchange of parts of Andy's medical record under controlled circumstances.
Sometimes they may need it very quickly, even if Andy is not in a conscious state. 
In the future, doctors prescribe recipes that \texttt{MyVirtualME} automatically transfers to the pharmacy of Andy’s choice, potentially selected based on availability and cost of the medicine.
\texttt{MyVirtualME} also organizes the transfer to Andy’s home.
Sometimes, the medication is already at Andy's house before Andy arrives. 
If Andy has defined a fully automated workflow for such cases, the bill can be automatically forwarded to Andy’s healthcare insurance, which then pays the pharmacy directly.
If healthcare insurance does not cover it completely, \texttt{MyVirtualME} transfers the needed difference from Andy’s bank account and submits a consideration proposal in the tax statement to his fiscal authorities, when appropriate.
No further processing is required by Andy.

The medical letters from his doctor are collected by \texttt{MyVirtualME} and forwarded to any of Andy’s doctors who are interested or should be aware of them. \texttt{MyVirtualME} also knows when and how to take the medicine to remind Andy and can handle further recommendations, such as showing up at the doctor's again in two weeks.

\subsection{Some more exemplaric scenarios}\label{sec:scenario2}

Similar scenarios can be identified in many other domains, e.g., in financial management, with banks, life insurance companies, brokers, crypto companies, and other financial organizations, where Andy is eager to control the flow of money and other assets through \texttt{MyVirtualME}.
If Andy pays for a business dinner or the fuel for a business trip, \texttt{MyVirtualME} directly forwards that to the employer because \texttt{MyVirtualME} knows from Andy's calendar that these expenses are related to his business trip.
As \texttt{MyVirtualME} filters and selects Andy's \textit{social media feed} according to his explicitly selected preferences, e.g., filter hate and otherwise negative comments, Andy regains his personal social sovereignty. Social media algorithms must adapt to align with Andy's preferences. 
Large projects, such as the \textit{construction of a new home}, involve many actors and their coordination, including obtaining permissions from various local and government administrative offices.
\texttt{MyVirtualME} knows which funding opportunities exist and which laws and other restrictions may apply.
In this way, it can make recommendations to promote the execution of the project or reduce costs.
\textit{Controlling Andy's smart home} becomes finally effective and even more energy-efficient.
Furthermore, as \textit{smart humanoid robots} emerge, they can better serve Andy's personal needs when deeply integrated with, or even completely controlled by \texttt{MyVirtualME}.

\section{Realizing a Trustworthy and Intelligent \texttt{MyVirtualME}}\label{sec:realisation}

\texttt{MyVirtualME} is not primarily about deploying a single technological artifact, but about establishing a socio-technical platform ecosystem in which humans, digital services, and institutions interact through a trusted, transparent, and evolvable digital representation of the individual. This perspective shifts the focus from isolated tools to the integration of services, the governance of data, and the embedding of intelligence in the digital infrastructure, so that human sovereignty is preserved.

\subsection{\texttt{MyVirtualME} as a Trusted and Intelligent Digital Twin}

At the center of this vision lies a logical platform that serves as a persistent, trustworthy digital twin of the individual~\cite{Bran25,VHT25} in the digital world. It hosts the internal services that make up \texttt{MyVirtualME}, integrates complementary roles into a single human-centered environment, maintains heterogeneous and context-dependent personal data over long periods, executes internal services that reason over those data, acts on behalf of the individual, and offers an intuitive and transparent interface for monitoring and control. At the same time, it exposes standardized, secure interfaces that allow external services to interact with \texttt{MyVirtualME} under explicitly defined, revocable conditions.

A fundamental design principle is the strict distinction between \textit{trusted internal services} and \textit{untrusted external services}. External services never gain unrestricted access to personal data. All exchanges are mediated by \texttt{MyVirtualME} trusted services, which enforce user-defined authorization policies, purpose limitations, and temporal constraints. Each interaction is logged with rich metadata that describes who requested which data, for what purpose, and under what conditions. This supports regulatory compliance while enabling a form of \textit{digital self-awareness}, allowing individuals to inspect, adapt, or revoke previously granted privileges at any time.

An important property of the platform is the use of \textit{bidirectional data links}, which allow relevant changes to propagate in both directions under user-defined policies. For example, updates to personal information can be selectively shared with trusted institutions, while changes to contractual terms, service conditions, or regulatory requirements are pushed back to \texttt{MyVirtualME} and surfaced to the user in a structured, actionable way. The digital interaction thus evolves from a sequence of isolated transactions into a continuous and transparent relationship.

\texttt{MyVirtualME} composes active services that allows humans to define rules for automating or enabling effective assistance with workflows (e.g., handling incoming requests, bills, data, ...). 
The rule-adaptive services are supported by trusted AI agents that operate within the trusted core, close to personal data, and interact with external services only through policy-governed, human-controlled, and sandboxed interfaces. 
These agents augment human decision-making and rule-definition by monitoring contexts, detecting relevant changes, and proposing actions that are consistent with user preferences, values, and legal constraints. 
In practice, rules and agents analyze the incoming information streams, identify obligations or opportunities, and manage workflows such as preparing documentation, coordinating appointments, or negotiating service conditions within predefined boundaries. 

\subsection{Interoperability, Standards, and Digital Sovereignty}

Trust in this infrastructure cannot be achieved if it is controlled by a single, external provider or authority. A core requirement is the existence of multiple independent \texttt{MyVirtualME} providers that adhere to shared technical and organizational standards. Individuals must be able to choose, migrate, and potentially federate their \texttt{MyVirtualME} across providers without losing control over their data or digital identity. This implies open and extensible standards for (1) initiating interactions between external stakeholders and \texttt{MyVirtualME}, (2) establishing and evolving bidirectional data connections, (3) expressing and (4) enforcing usage conditions such as purpose, scope, and retention. It also requires mechanisms to (5) propagate revocation of previously granted access rights and to (6) discover relevant services based on user-defined goals and contexts.

A prerequisite in this setting is a \textit{unique, persistent, and user-controlled digital identifier} that enables secure and privacy-preserving linkage of personal data across services. Unlike today’s email or platform-bound identifiers, this identifier is owned by the individual and is managed through \texttt{MyVirtualME}, ensuring continuity and independence from specific service providers.

The broader implications extend beyond improving individual interactions with existing services. If organizations expose their digital processes through standardized interfaces to \texttt{MyVirtualME}, new services become accessible via configuration rather than requiring repeated registration, manual data entry, or ad hoc coordination. The digital transformation can then accumulate systematically without increasing the cognitive or administrative burden on individuals.  
Revisiting common examples of inefficient digitalization (cf. Sections~\ref{sec:scenario1} and~\ref{sec:scenario2}), \texttt{MyVirtualME} enables fundamentally different workflows. Once a document or data item exists within the trusted core, it can be selectively and transparently reused across institutions under user-defined policies, with full traceability and the possibility of revocation. What is currently experienced as fragmentation and redundancy is transformed into coherent human-centered digital choreography.

\section{A Call for Action}\label{subsec:call}

Developing \texttt{MyVirtualME} is a major interdisciplinary challenge at the intersection of, e.g., software engineering, distributed systems, data science, AI, security and privacy engineering, and human-computer interaction. 
Although many elements already exist~\cite{Poi15,PO17,Conversation25}, putting it all together is an undertaking. 
Key open questions include scalable architectures for long-term personal data management services, precise models for purpose- and policy-aware data exchange, verifiable sandboxing of untrusted services, and explainable AI agents. 
From a societal perspective, this vision offers a pathway toward a digitally integrated society that strengthens, rather than weakens, individual autonomy. 
It enforces transparency, bidirectionality of data/knowledge, and user-controlled intelligence at the core of digital interaction, and thus \texttt{MyVirtualME} provides a solid foundation for balancing innovation and fundamental rights in large-scale digital ecosystems. 
In this sense, the establishment of \texttt{MyVirtualME} is not merely a technical endeavor, but a strategic investment in a trustworthy, human-centered digital future. 

This is a \textbf{call for action!} Let us start designing and developing such a platform in a reliable, scalable, and trustworthy way. Let us act fast and not discuss for too long.

\bibliographystyle{plain}
\bibliography{references,sselit}

\end{document}